\documentstyle[12pt]{article}
 \title{Relativistic Spectral  Properties of Landau Operator with $\delta-$ and $\delta '-$ cylinder interactions}

\author{G.~Honnouvo and M. N.~Hounkonnou}
\date{}

\catcode `\@=11
\@addtoreset{equation}{section}

\newtheorem{thm}{Theorem}[section]
\newif\if@cghi
\def\cite{\@cghitrue\@ifnextchar [{\@tempswatrue
	\@citex}{\@tempswafalse\@citex[]}}
\def\citelow{\@cghifalse\@ifnextchar [{\@tempswatrue
	\@citex}{\@tempswafalse\@citex[]}}
\def\@cite#1#2{{$^{#1}$\if@tempswa\typeout
	{IJCGA warning: optional citation argument	ignored: `#2'} \fi}}

 \def \R {{\rm I \! R }}
\def \C {{\rm l \! \! \! C }}

\def \Z {{\rm Z \! \! Z }}

\begin{document}
\maketitle

\begin{quote}
{\small \quad Unit\'e de Recherche en Physique Th\'eorique (URPT)\\
\phantom{e)x}Institut de Math\'ematiques
et de Sciences  Physiques (IMSP) \\
\phantom{e)x}01 B.P. 2628 Porto-Novo, BENIN, \\
 \phantom{e)x}International Chair in Mathematical Physics and Applications (ICMPA)\\
\phantom{e)x}01 BP 2628 Porto-Novo, BENIN,\\
 \rm \phantom{e)x}g$_{-}$honnouvo@yahoo.fr,
 hounkonnou1@yahoo.fr}
\vspace{8mm}

\noindent {\small \begin{abstract}Using the theory of self-adjoint extensions, we study the relativistic  spectral properties of the Landau operator with $\delta$ and $\delta '$ interactions on a cylinder of radius $R$ for a charged spin particle system, formally given by the Hamiltonian 
$H^G_{B} = {(p-A)}^2 1\!\!1 + {\bf{\sigma}}.{\bf{B}+G}V(r),$ $\bigg(V(r) = \delta (R-r)$ or $V(r) = \delta' (R-r) \bigg),$
acting in $ L^2(\R^2)\bigotimes \C^2  $. $G$  a scalar $2\times 2$ real matrix. $1\!\!1$ is the identity matrix. The potential vector has the form $A=
(B/2)(-y,\: x   )$ and $B>0.$

\end{abstract}}
\end{quote}

\section{Introduction}
Over the last decades, there have been major research efforts in studying Schr\"odinger operators to describe properties of charged particles in magnetic systems (see \cite{two}, \cite{Pa} and references therein). Gesztesy et al.\cite{GHS} have devoted a communication to the point interaction in magnetic field systems in the nonrelativistic case . More recently \cite{HH1}, we have studied the nonrelativistic case of this model.
In this paper, we deal with the relativistic models
 for a two-dimensional quantum Hamiltonian describing a charged spin particle in
  a constant and uniform magnetic field superimposed to  $\delta$ and $\delta '$-interaction
   on a cylinder of radius $R.$

We consider an electron confined in the $x-y$ plane and subjected to a uniform magnetic
 field perpendicular to the plane, using the symmetric gauge vector potential $A= (B/2)(-y,\: x   ),\: B > 0.$ 

We provide a complete spectral analysis of the given operator
and deduce helpful properties when this operator is perturbed by $\delta$ or $\delta '$
 interaction on the cylinder of radius $R.$ We recover the results obtained in \cite{GHS}, \cite{HH1}
  for the point interaction as a particular case of our study.

First, using the von Neumann theory of self-adjoint $(s.a.)$ extension of linear symmetric operators \cite{AG}, we characterized the self-adjoint extension of this operator. The relativistic  type Hamiltonian is formally expressed as:

\begin{equation}\label{a4} H^G_{B}={(p-A)}^2 1\!\!1 +
\left( \begin{array}{ll}\nonumber
 B +\alpha V(r)\qquad \qquad&  0\\\\0 & - B +\beta V(r)\end{array}\nonumber
\right).\nonumber
\end{equation}
where
\begin{equation}\label{a3}
V(r) = \left | \begin{array}{ll}
\xi \delta (r-R), \\
\xi \delta '(r-R), \\
\end{array}\right.\\
\\
\:\:\:\:\:\mbox{with} \:\: \xi\in \R,\:\: R>0.
\end{equation}

The paper is organized as follows: In section 2, we investigate
 the relativistic model of $\delta$.  In section 3, the relativistic model of $\delta'$. Finally, in section 4, we conclude
with some remarks.

\section{The model of the relativistic $\delta $-cylinder interaction}

In this section, we derive the properties of the relativistic quantum Hamiltonian describing a charged spin particle in a constant magnetic field ${\bf{B}}$ coupled with $\delta$ interaction on the cylinder of radius $R,$ given by

\begin{eqnarray}\label{c1}
H^G_{B} = {(p-A)}^2  1\!\!1 + {\bf{\sigma}}.{\bf{B}+G}\delta (R-r),
\end{eqnarray}
where

\begin{eqnarray}\label{c2}
{\bf{\sigma}} =(\sigma^1, \sigma^2, \sigma^3),
\end{eqnarray}
and the $\sigma^i$ are the Pauli matrices defined by

\begin{equation}\label{c3}
 \sigma^1=
\left( \begin{array}{ll}
0\:\:\:\: & 1\\\\1 & 0 \end{array}
\right),
\quad \sigma^2=
\left( \begin{array}{ll}
0 \:\:\:\: & -i\\\\i & 0 \end{array}
\right),
\quad \sigma^3=
\left( \begin{array}{ll}
1 \:\:\:\: & 0\\\\0 & -1 \end{array}
\right)\mbox {and}
\quad G=
\left( \begin{array}{ll}
\alpha \:\:\:\: & 0\\\\0 & \beta \end{array}
\right).
\end{equation}
$\alpha$ and $\beta$ are real numbers.

Let us consider the operator

\begin{eqnarray}\label{c4}
H_{B} = {(p-A)}^2  1\!\!1 + \bf{\sigma}.\bf{B},
\end{eqnarray}

and the closed symmetric operator

\begin{equation}\label{c5} \dot H_{B}=
\left( \begin{array}{cl}
{(p-A)}^2 +B& \qquad \qquad  \qquad 0 \\\\
0 & {(p-A)}^2 -B\end{array}
\right),
\end{equation}

with the domain
\begin{eqnarray}\label{c6}
{\cal{D}}(\dot{H_B})=\{\psi\in H^{2,2}(\R^2)\otimes \C^2 \;,
\; \psi(S^{R})=0 ; \dot H_B \psi \in L^2(\R^2)\otimes \C^2 \}\;,
\end{eqnarray}
where
$S^{R} =\{\underline{x}\in \R^2 \;,\; |\underline{x}|=R\}$ is a circle of radius $R$ centered at the  origin in $\R^2$,
and $H^{k,p}(\Omega)$ is the Sobolev space of indices $(k,p).$

\noindent

Let us now decompose the Hilbert space ${\cal H}=L^2(\R^2)\bigotimes \C^2$ as follows

\begin{equation}\label{c7}
{\cal H}=L^2(\R^2)\bigotimes \C^2 = \bigg(L^2(]0, \infty[)\bigotimes L^2(S^1) \bigg) \bigotimes \C^2,
\end{equation}
$S^1$ being the unit circle in $\R^2$.

 Similarly to the nonrelativistic study, the following isometry is introduced in order to remove the weight factor $r$ from the measure:

\begin{equation}\label{c8} \tilde U:\left\{ \begin{array}{c} L^2((0,\infty );rdr)\longrightarrow L^2((0,\infty
);dr)\equiv L^2((0,\infty ))\\ f\longmapsto \left( \tilde Uf\right) (r)=\sqrt{r}f(r).\end{array} \right.
\end{equation}
Then, we get the following decomposition of the Hilbert space ${\cal{H}}$:\\

\begin{equation}\label{c9}
{\cal{H}} = \bigoplus_{m'={-\infty}}^{m'= +\infty}\bigoplus_{m={-\infty}}^{m= +\infty}\Bigg [\bigg (\tilde U^{-1}(L^2(]0; \infty))\bigotimes \left[\frac{e^{im\phi}}{\sqrt{2\pi}} \right]\bigg )\bigoplus \bigg (\tilde U^{-1}(L^2(]0; \infty))\bigotimes \left[\frac{e^{im'\theta}}{\sqrt{2\pi}} \right]\bigg )  \Bigg ],
\end{equation}

so that the operator $\dot{H_B}$  writes
\begin{equation}\label{c10}
{\dot H_B} = \bigoplus_{m= -\infty}^{m= +\infty}\bigoplus_{m'= -\infty}^{m'= +\infty}\;
 \tilde U^{-1}{\dot H}_{B,m,m'}\tilde U
\bigotimes{1\!\!1 },
\end{equation}

where the radial part $\dot H_{B,m,m'}$ is defined by
\begin{eqnarray}\label{c11}\dot H_{B,m,m'}=
\left( \begin{array}{cl}
h_{B,m}^1 & \quad 0\\\\
0 &h_{B,m'}^2\end{array}
\right),
\end{eqnarray}

$$h_{B,m}^1 = -\frac{d^2}{dr^2}+{\bigg(\frac{ m }{r}+ \frac{B}{2}r\bigg)}^2- \frac{1}{4r^2} + B$$ and
$$h_{B,m'}^2 = -\frac{d^2}{dr^2}+{\bigg(\frac{ m' }{r}+ \frac{B}{2}r\bigg)}^2- \frac{1}{4r^2} - B$$

with the domain

\begin{eqnarray}\label{c13}
{\cal D}(\dot{H}_{B,m,m'}) & = & \left\{ f\in \bigg (L^2(]0, \infty [, dr)\cap H^{2,2}_{loc}(]0, \infty [)\bigg) \bigotimes \C^2\, ;\right.
 f(R_{\pm})=0\, ;\nonumber \\
     &     &  \left.  \dot H_{B,m,m'}f \in L^2((0,\infty ))\bigotimes \C^2\:\right\},\:\: m\in {\Z},\: m'\in {\Z}.
\end{eqnarray}

The adjoint operator $\dot H^*_{B , m,m'}$ of $\dot H_{B , m,m'}$ is defined by (\ref{c11})
in the domain
\begin{eqnarray}\label{c15}
{\cal D}(\dot{H}^*_{B,m,m'}) & = & \left\{ f\in \bigg (L^2(]0, \infty [, dr)\cap H^{2,2}_{loc}(]0, \infty [\setminus\{R\})\bigg) \bigotimes \C^2\, ;\right.
 f(R_+)=f(R_-)\equiv f(R);\nonumber \\
& & \left. \dot H_{B,m,m'}f \in L^2((0,\infty ))\bigotimes \C^2\, \,   \right\},\nonumber \\
& & m\in {\ Z},\: m'\in {\Z}
\end{eqnarray}

The indicial equation reads

$$\dot H^*_{B , m,m'}\tilde \psi_{B,m,m'}(k,r) = k\tilde \psi_{B,m,m'}(k,r),$$  \\or equivalently

\begin{equation}\label{c16} \dot H^*_{B,m,m'}
\left(\begin{array}{l} \phi_{B,m,m'}^1\\ \phi_{B,m,m'}^2
\end{array}\right)  = k \left(\begin{array}{l} \phi_{B,m,m'}^1\\ \phi_{B,m,m'}^2
\end{array}\right),
\end{equation}
$k\in \C\setminus \R,$ which has two solutions:

\begin{equation}\label{c17} \tilde \psi_{B,m,m'}^1 =
\left(\begin{array}{l} \phi_{B,m,m'}^1\\ 0
\end{array}\right)
\end{equation}
and
\begin{equation}\label{c18} \tilde \psi_{B,m,m'}^2 =
\left(\begin{array}{l} 0\\ \phi_{B,m,m'}^2
\end{array}\right)
\end{equation}

where

\begin{equation}\label{c19}
\phi_{B, m,m'}^1(k,r)  = \left\{\begin{array}{ll}
N_{B,m}^1G^{(1)}_{B,m}(k,R)\times
F_{B, m}^{(1)}(k,r)\;;           &r \leq R\;,\\
N_{B,m}^1F^{(1)}_{B, m}(k,R)\times
G_{B, m}^{(1)}(k,r)\;;       &r \geq R\;,
\end{array}\right.
\end{equation}

\begin{equation}\label{c20}
\phi_{B,m, m'}^2(k,r)  = \left\{\begin{array}{ll}
N_{B,m'}^2 G^{(2)}_{B,m'}(k,R)\times
F_{B, m'}^{(2)}(k,r)\;;           &r \leq R\;,\\
N_{B,m'}^2 F^{(2)}_{B, m'}(k,R)\times
G_{B, m'}^{(2)}(k,r)\;;       &r \geq R\;,
\end{array}\right.
\end{equation}

\begin{eqnarray}\label{c21}
\begin{array}{ll}
F_{B, m}^{(1)}(k,r) = {r}^{ 1/2 + |m|}e^{-\frac{1}{4}Br^2} {}_1F_1 \left({1\over 2} (|m| + m +1 -\frac{k-B}{B}), |m| +1; \frac{B}{2}r^2 \right),\\\\
G_{B, m}^{(1)}(k,r) ={r}^{ 1/2 +|m|} e^{-\frac{1}{4}Br^2} U \left({1\over 2} (|m| + m +1 -\frac{k-B}{B}), |m| +1; \frac{B}{2}r^2 \right),
\end{array}
\end{eqnarray}

\begin{eqnarray}\label{c22}
\begin{array}{ll}
F_{B, m'}^{(2)}(k,r) = {r}^{ 1/2 + |m'|}e^{-\frac{1}{4}Br^2} {}_1F_1 \left({1\over 2} (|m'| + m' +1 -\frac{k+B}{B}), |m'| +1; \frac{B}{2}r^2 \right),\\\\
G_{B, m'}^{(2)}(k,r) ={r}^{ 1/2 +|m'|} e^{-\frac{1}{4}Br^2} U \left({1\over 2} (|m'| + m' +1 -\frac{k+B}{B}), |m'| +1; \frac{B}{2}r^2 \right),
\end{array}
\end{eqnarray}

\begin{eqnarray}\label{c23}
N_{B, m}^1 = \bigg ( ||P_{B, m}^1 (k)||_{L^2 (]0, \infty))} \bigg )^{-1},
\end{eqnarray}

\begin{eqnarray}\label{c23O}
N_{B, m'}^2 = \bigg ( ||P_{B, m'}^2 (k)||_{L^2 (]0, \infty))} \bigg )^{-1},
\end{eqnarray}

\begin{equation}\label{c24}
P_{B, m}^1(k,r)  = \left\{\begin{array}{ll}
G^{(1)}_{B,m}(k,R)\times
F_{B, m}^{(1)}(k,r)\;;           &r \leq R\;,\\
 F^{(1)}_{B, m}(k,R)\times
G_{B, m}^{(1)}(k,r)\;;       &r \geq R\;,
\end{array}\right.
\end{equation}

and

\begin{equation}\label{c25}
P_{B, m'}^2(k,r)  = \left\{\begin{array}{ll}
G^{(2)}_{B,m'}(k,R)\times
F_{B, m'}^{(2)}(k,r)\;;           &r \leq R\;,\\
 F^{(2)}_{B, m'}(k,R)\times
G_{B, m'}^{(2)}(k,r)\;;       &r \geq R\;.
\end{array}\right.
\end{equation}

Since the indicial equation admits two solutions, $\dot H_{B,m,m'}$ has
deficiency indices $(2,2)$  and, consequently, all self-adjoint (s.a)
extensions of $\dot H_{B , m,m'}$ are given by a $4$-parameter family of (s.a.)
operators\cite{AG}. In particular, we define here

\begin{eqnarray}\label{c26} H^{G_{m,m'}}_{B,m,m'}=
\left( \begin{array}{cl}
h_{B,m}^1 & \quad 0\\\\
0 &h_{B,m'}^2\end{array}
\right)
\end{eqnarray}

with the domain
\begin{eqnarray}\label{c27}
{\cal D}({H}^{G_{m,m'}}_{B, m,m'}) & = & \left\{ f\in \bigg (L^2(]0, \infty [, dr)\cap H^{2,2}_{loc}(]0, \infty [-\{R\})\bigg) \bigotimes \C^2\, ;\right.
 f(R_+)=f(R_-)\equiv f(R);\nonumber \\
& &f'(R_+)-f'(R_-) = G_{m,m'} f(R); \left.  H^{G_{m,m'}}_{B,m,m'}f \in L^2((0,\infty ))\bigotimes \C^2\, \,   \right\},\nonumber \\
& & m\in {\Z},\: m'\in {\Z}
\end{eqnarray}

and

\begin{equation}\label{c28}
\quad G_{m,m'}=
\left( \begin{array}{ll}
\alpha_m \:\:\:\: & 0\\\\0 & \beta_{m'} \end{array}
\right).
\end{equation}

The case $G_{m,m'} = 0$ coincides with the free kinetic energy Hamiltonian $ H^{O}_{B, m,m'}$  for fixed quantum numbers $m, m'.$

Let $G=\{ G_{m,m'} \}_{m, \:m'\in \Z}$ and introduce in $L^2(\R^2) \bigotimes \C^2$ the operator

\begin{equation}\label{c29}
  H^{G}_{B}= \bigoplus_{m={-\infty}}^{m= +\infty}\bigoplus_{m'={-\infty}}^{m'= +\infty}\tilde U^{-1} H^{G_{m,m'}}_{B , m,m'}\tilde U\bigotimes 1\!\!1.
\end{equation}
By definition, $H^{G}_{B}$ is the rigorous mathematical formulation of the formal expression $(\ref{c1}).$\\
Actually, it provides a slight generalization of $(\ref{c1})$, since $G$
may depend on $m\in \Z$ and $m'\in \Z.$

\smallskip\begin{thm}\label{2.1}
(i) The resolvent of $ H^{G_{m,m'}}_{B, m,m'}$ is given by

\begin{eqnarray}\label{c30}
{ (  H^{G_{m,m'}}_{B, m,m'} -k)}^{-1} = { ( H^{O}_{B, m,m'} -k)}^{-1} + \sum_{i,j =1}^{2}\mu_{i,j}(k)\left(\tilde \Psi_{B, m,m'}^i(\overline{k}), .\right)\tilde \Psi_{B, m,m'}^j(k),\nonumber \nonumber
\end{eqnarray}
\begin{equation}\label{20}
k\in \rho ( H^{G_{m,m'}}_{B,m,m'})\cap\rho ( H^{O}_{B,m,m'}) ,\:\:m\in\Z,\: m'\in\Z,
\end{equation}
where
\begin{equation}\label{c31}
\mu_{1,1}(k) = \frac{\alpha_m}{N^{1}_{B,m}\bigg(G'^{(1)}_{B, m}(k,R)F^{(1)}_{B, m}(k,R)
 -G^{(1)}_{B, m}(k,R)F'^{(1)}_{B, m}(k,R) - \alpha_m G^{(1)}_{B, m}(k,R)F^{(1)}_{B, m}(k,R) \bigg)},
\end{equation}

\begin{equation}\label{c32}
\mu_{1,2}(k) = 0,
\end{equation}

\begin{equation}\label{c33}
\mu_{2,1}(k) = 0
\end{equation}

and
\begin{equation}\label{c34}
\mu_{2,2}(k) = \frac{\beta_{m'}}{N^{2}_{B,m'}\bigg(G'^{(2)}_{B, m'}(k,R)F^{(2)}_{B, m'}(k,R) -G^{(2)}_{B, m'}(k,R)F'^{(2)}_{B, m'}(k,R) - \beta_{m'} G^{(2)}_{B, m'}(k,R)F^{(2)}_{B, m'}(k,R) \bigg)}.
\end{equation}

The Green function $G_{B,m,m'}^O(k,r,r')$ of $H^{O}_{B,m,m' } $ has the form

\begin{equation}\label{c35} {(H^{O}_{B,m,m' } -k)}^{-1}= G_{B,m,m'}^O(k,r,r')=
\left( \begin{array}{ll}
g^{1,1}_{m,m'}(k,r,r')\qquad \qquad  \qquad& g^{1,2}_{m,m'}(k,r,r')\\\\   g^{2,1}_{m,m'}(k,r,r') \qquad \qquad  \qquad&  g^{2,2}_{m,m'}(k,r,r')\end{array}
\right),
\end{equation}
where

\begin{eqnarray}\label{c36}
g^{1,2}_{m,m'}(k,r,r')(r,r') = g^{2,1}_{m,m'}(k,r,r')=0 ,
\end{eqnarray}

\begin{equation}\label{c37}
g^{1,1}_{m,m'}(k,r,r')= \left\{\begin{array}{ll}
N^{1}_{B,m}G^{(1)}_{B, m}(k,r)\times
F_{B,m}^{(1)}(k,r')\;;           &r' \leq r\;,\\
 N^{1}_{B,m}F^{(1)}_{B,m}(k,r)\times
G_{B,m}^{(1)}(k,r')\;;       &r' \geq r\;
\end{array}\right .
\end{equation}

and

\begin{equation}\label{c38}
g^{2,2}_{m,m'}(k,r,r')= \left\{\begin{array}{ll}
N^{2}_{B,m'}G^{(2)}_{B, m'}(k,r)\times
F_{B,m'}^{(2)}(k,r')\;;           &r' \leq r\;,\\
 N^{2}_{B,m'}F^{(2)}_{B,m'}(k,r)\times
G_{B,m'}^{(2)}(k,r')\;;       &r' \geq r\;.
\end{array}\right.
\end{equation}

We note that $g^{1,1}_{m,m'}(k,R,r)= \phi_{B, m}^1(k,r)$ and $g^{2,2}_{m,m'}(k,R,r)= \phi_{B, m}^2(k,r).$ \\
(ii) The resolvent of $H^{G}_{B}$ is given by

\begin{eqnarray}\label{c39}
{\left( H_{B}^G  - k\right)}^{-1}= {\left( H_{B}^O  - k\right)}^{-1} + \bigoplus_{m=-\infty}^{m=+\infty}  \bigoplus_{m'=-\infty}^{m'=+\infty} \sum_{i,j =1}^{2}\mu_{i,j}(k)\left(|.|^{-1}\tilde \Psi_{B, m,m'}^i(\overline{k}), .\right)|.|^{-1}\tilde \Psi_{B, m,m'}^j(k),\nonumber \nonumber
\end{eqnarray}
\begin{equation}\label{22}
k\in \rho ( H^{G}_{B})\cap\rho ( H^{O}_{B}) , \:\: m\in\Z,\: m'\in\Z.
\end{equation}
\end{thm}\smallskip

{\it{Proof:}} Since $\dot H_{B,m,m'}$ has deficiency indices (2.2), it follows from Krein's formula \cite{AGHS}  that the resolvent of $ H^{G_{m,m'}}_{B,m,m'}$ is given by

\begin{eqnarray}\label{c40}
{ ( H^{G_{m,m'}}_{B , m,m'} -k)}^{-1} = { ( H^{0}_{B , m,m'} -k)}^{-1} + \sum_{i,j =1}^{2}\mu_{i,j}(k)\left(\tilde \Psi_{B, m,m'}^i(\overline{k}), .\right)\tilde \Psi_{B, m,m'}^j(k),\nonumber \nonumber
\end{eqnarray}
\begin{equation}\label{23}
k\in \rho ( H^{G_{m,m'}}_{B,m,m'})\cap\rho ( H^{O}_{B,m,m'}),  \:\:m\in\Z,\:m'\in\Z.
\end{equation}
Since $G_{B,m,m' }^{G_{m,m'}}(k,r,r')$ must satisfy the following equation:

\begin{equation}\label{c41} (H^{O}_{B,m,m' } -k)G_{B,m,m' }^{G_{m,m'}}(k,r,r')=
\left( \begin{array}{ll}
\delta(r-r')\qquad \qquad  \qquad& 0\\\\  \qquad 0 &  \delta(r-r')\end{array}
\right),
\end{equation}

one has

\begin{eqnarray}\label{c42}
{(h^1_{B,m} -z)}g^{1,1}_{m,m'}(k,r,r') =\delta(r-r'),
\end{eqnarray}

\begin{eqnarray}\label{c43}
g^{1,2}_{m,m'}(k,r,r') = g^{2,1}_{m,m'}(k,r,r')=0 ,
\end{eqnarray}

\begin{eqnarray}\label{c44}
{(h_{B, m'}^2 -z)}g^{2,2}_{m,m'}(k,r,r')=\delta(r-r').
\end{eqnarray}

which implies that

\begin{equation}\label{c45}
g^{1,1}_{m,m'}(k,r,r')= \left\{\begin{array}{ll}
N_{B, m}^1G^{(1)}_{B, m}(k,r)\times
F_{B,m}^{(1)}(k,r')\;;           &r' \leq r\;,\\
N_{B, m}^1 F^{(1)}_{B,m}(k,r)\times
G_{B,m}^{(1)}(k,r')\;;       &r' \geq r
\end{array}\right .
\end{equation}

and

\begin{equation}\label{c46}
g^{2,2}_{m,m'}(k,r,r')= \left\{\begin{array}{ll}
N_{B, m'}^2G^{(2)}_{B, m'}(k,r)\times
F_{B,m'}^{(2)}(k,r')\;;           &r' \leq r\;,\\
 N_{B, m'}^2F^{(2)}_{B,m'}(k,r)\times
G_{B,m'}^{(2)}(k,r')\;;       &r' \geq r\;.
\end{array}\right.
\end{equation}

For the determination of $\mu_{i,j}(k)$, we proceed as follows. Let $g\in L^2(]0,\infty])$ and define the function
 $$\chi_{m,m'}(k,r) = \left( {(  H^{G_{m,m'}}_{B , m,m'} -k)}^{-1}g\right) (r).$$
Since $\chi_{m,m'} \in D( H^{G_{m,m'}}_{B , m,m'})$, it follows that $\chi_{m,m'}$ should satisfy the boundary conditions of $D(H_{B , m, m'}^{G_{m,m'}})$, the implementation of which  gives

\begin{equation}\label{c47}
\mu_{1,1}(k) = \frac{\alpha_{m}}{N^{1}_{B,m}\bigg(G'^{(1)}_{B, m}(k,R)F^{(1)}_{B, m}(k,R)
-G^{(1)}_{B, m}(k,R)F'^{(1)}_{B, m}(k,R) - \alpha_m G^{(1)}_{B, m}(k,R)F^{(1)}_{B, m}(k,R) \bigg)},
\end{equation}

\begin{equation}\label{c48}
\mu_{1,2}(k) = 0,
\end{equation}

\begin{equation}\label{c49}
\mu_{2,1}(k) = 0
\end{equation}

and
\begin{equation}\label{c50}
\mu_{2,2}(k) = \frac{\beta_{m'}}{N^{2}_{B,m'}\bigg(G'^{(2)}_{B, m'}(k,R)F^{(2)}_{B, m'}(k,R) -G^{(2)}_{B, m'}(k,R)F'^{(2)}_{B, m'}(k,R) - \beta_{m'} G^{(2)}_{B, m'}(k,R)F^{(2)}_{B, m'}(k,R) \bigg)}.
\end{equation}

Inserting (\ref{c47}), (\ref{c48}) (\ref{c49}) and (\ref{c50})   into ((\ref{c40})), we deduce the expression  (\ref{c30}).
Eq. ((\ref{c39}) follows from (\ref{c29}) and ((\ref{c30}) $.$\\

Spectral properties of $ H^{G_{m,m'}}_{B ,m,m'}$ are provided by the following theorem where
$\sigma_{ess}(.)\, , \, \sigma_{sc}(.)$ and
$\sigma_p(.)$ represent the same properties as mentioned.

\smallskip\begin{thm}\label{2.3}:
For all $\alpha_m\in (-\infty ,\infty )$ and $\beta_{m'}\in (-\infty ,\infty ),$ we have the following
results
\begin{eqnarray}\label{sppp}
\sigma_{ess}( H^{G_{m,m'}}_{B ,m,m'}) = \emptyset,
\end{eqnarray}

\begin{eqnarray}\label{s}
\sigma_{sc}( H^{G_{m,m'}}_{B ,m,m'})= \emptyset,
\end{eqnarray}
\begin{eqnarray}\label{sp24}
\sigma_p(H^{G_{m,m'}}_{B ,m,m'}) & =&\bigg \{E\in\R / G'^{(1)}_{B, m}(E,R)F^{(1)}_{B, m}(E,R) -G^{(1)}_{B, m}(E,R)F'^{(1)}_{B, m}(E,R)\nonumber \\
& &- \alpha_m G^{(1)}_{B, m}(E,R)F^{(1)}_{B, m}(E,R) =0 \nonumber\\
& & \mbox{or} \: \: G'^{(2)}_{B, m'}(E,R)F^{(2)}_{B, m'}(E,R) -G^{(2)}_{B, m'}(E,R)F'^{(2)}_{B, m'}(E,R)\nonumber \\
& & - \beta_{m'} G^{(2)}_{B, m'}(E,R)F^{(2)}_{B, m'}(E,R) =0    \bigg\}.
\end{eqnarray}

The negative bound states are related to the  eigenvalues of $ H^{G_{m,m'}}_{B ,m,m'}$ obtained from the equation
\begin{equation}\label{c'40}
{det(\mu_{i,j}(E))}^{-1}= 0;\:\:E<0,
\end{equation}
which has at most two solutions $E_0 <0.$
\end{thm}\smallskip

{\it{Proof: }}
We know that \cite{two} $\sigma_{ess}(H^{O}_{B ,m,m'})= \emptyset.$ Using Weyl's theorem \cite{RS}, we have
\begin{equation}
\sigma_{ess}(H^{G_{m,m'}}_{B ,m,m'})=\sigma_{ess}(H^{O}_{B ,m,m'})= \emptyset.
\end{equation}
Eq.(\ref{c39}) implies that the point spectrum is determined by

\begin{eqnarray}
\sigma_p(H^{G_{m,m'}}_{B ,m,m'}) &=& \bigg \{ E\in \R / \: {det(\mu_{i,j}(E))}^{-1}=0 \bigg\} \nonumber \\
& =& \bigg \{ E\in \R / \: G'^{(1)}_{B, m}(E,R)F^{(1)}_{B, m}(E,R) -G^{(1)}_{B, m}(E,R)F'^{(1)}_{B, m}(E,R)\nonumber\\
& & - \alpha_m G^{(1)}_{B, m}(E,R)F^{(1)}_{B, m}(E,R) =0 \nonumber\\
 & &\mbox{or}\:  G'^{(2)}_{B, m'}(E,R)F^{(2)}_{B, m'}(E,R) -G^{(2)}_{B, m'}(E,R)F'^{(2)}_{B, m'}(E,R)\nonumber\\
 & & - \beta_{m'} G^{(2)}_{B, m'}(E,R)F^{(2)}_{B, m'}(E,R) =0    \bigg\}\nonumber,
\end{eqnarray}
which proves $(\ref{sp24}).$
The last part of this theorem follows from the statement of\cite{AG} [theorem 1, page 116] $.$

\section{ The model of relativistic $\delta' $- cylinder interaction}

In this section, we deal with the relativistic quantum Hamiltonian of a charged spin particle in a constant  uniform magnetic field ${\bf{B}}$ coupled with $\delta '$ interaction on the cylinder of radius $R,$ given by

\begin{eqnarray}\label{d1}
H^G_{B} = {(p-A)}^2  1\!\!1 + {\bf{\sigma}}.{\bf{B}+G}\delta '(R-r).
\end{eqnarray}

Let us consider the closed symmetric operator, formally defined by
\begin{equation}
\dot{H_B} ={(p-A)}^2  1\!\!1 + {\bf{\sigma}}.{\bf{B}},
\end{equation}

with the domain
\begin{eqnarray}
{\cal{D}}(\dot{H_B})=\{\psi\in H^{2,2}(\R^2)\otimes \C^2 \;,
\; \psi '(S^{R})=0 ; \dot H_B \psi \in L^2(\R^2)\otimes \C^2 \}\;.
\end{eqnarray}

Performing the same decomposition (\ref{c9}) and (\ref{c10}) for the Hilbert
 space ${\cal{H}}$  and for the operator $\dot H_B,$ we readily recover
  Eq.(\ref{c11}) for the operator $\dot H_{B,m,m'}$:

\begin{eqnarray}\label{122}\dot H_{B,m,m'}=
\left( \begin{array}{cl}
h_{B,m}^1 & \quad 0\\\\
0 &h_{B,m'}^2\end{array}
\right)
\end{eqnarray}

with

\begin{eqnarray}\label{d10}
{\cal D}(\dot H_{B,m,m'}) & = & \left\{ f\in \bigg (L^2(]0, \infty [, dr)\cap H^{2,2}_{loc}(]0, \infty [)\bigg) \bigotimes \C^2\, ;\right.
 f'(R_{\pm})=0\, ;\nonumber \\
     &     &  \left. \dot H_{B,m,m'}f \in L^2((0,\infty ))\bigotimes \C^2\, \,   \right\},\:\: m,\:m'\in {\Z}.
\end{eqnarray}
Here the boundary conditions require the derivative $f'$
to vanish at the circle of radius $R$ for the $\delta '$-interaction instead of the continuity of the function $f$ at $r=R$ in the case of $\delta$-interaction (see Eq. (\ref{c13})).

The adjoint operator $\dot H^*_{B , m,m'}$ of $\dot H_{B , m,m'}$ is defined by

\begin{eqnarray}\label{122}\dot H^*_{B,m,m'}=
\left( \begin{array}{cl}
h_{B,m}^1 & \quad 0\\\\
0 &h_{B,m'}^2\end{array}
\right)
\end{eqnarray}

with the domain

\begin{eqnarray}\label{d12}
{\cal D}(\dot{H}^*_{B,m}) & = & \left\{ f\in \bigg (L^2(]0, \infty [, dr)\cap H^{2,2}_{loc}(]0, \infty [\setminus\{R\})\bigg) \bigotimes \C^2\, ;\right.
 f'(R_+)=f'(R_-)\equiv f'(R)\, ;\nonumber \\
     &     &  \left. \dot H_{B,m,m'}f \in L^2((0,\infty ))\bigotimes \C^2\, \,   \right\},\:\: m,\:m'\in {\Z}.
\end{eqnarray}
The indicial equation reads

$$\dot H^*_{B , m,m'}\tilde \psi^1_{B,m,m'}(k,r) = k \tilde \psi^1_{B,m,m'}(k,r),$$  \\or equivalently

\begin{equation}\label{d13} \dot H^*_{B,m,m'}
\left(\begin{array}{l} \phi^1_{B,m}\\ \phi^2_{B,m'}
\end{array}\right)  = k \left(\begin{array}{l} \phi^1_{B,m}\\ \phi^2_{B,m'}
\end{array}\right),
\end{equation}
$k\in \C\setminus \R,$ which has two solutions:

\begin{equation}\label{d14} \tilde \psi^1 =
\left(\begin{array}{l} \phi^1_{B,m}\\ 0
\end{array}\right)
\end{equation}
and
\begin{equation}\label{d15} \tilde \psi^2 =
\left(\begin{array}{l} 0\\ \phi^2_{B,m'}
\end{array}\right)
\end{equation}

where

\begin{equation}\label{d16}
\phi_{B, m}^1(k,r)  = \left\{\begin{array}{ll}
M_{B,m}^1\big [G^{(1)}_{B,m}(k,r)\big]'_{r=R}\times
F_{B, m}^{(1)}(k,r)\;;           &r < R\;,\\
M_{B,m}^1 \big[F^{(1)}_{B, m}(k, r)\big]'_{r=R}\times
G_{B, m}^{(1)}(k,r)\;;       &r > R\;,
\end{array}\right.
\end{equation}

\begin{equation}\label{d17}
\phi_{B, m'}^2(k,r)  = \left\{\begin{array}{ll}
M_{B,m'}^2 \big[G^{(2)}_{B,m'}(k,r)\big]'_{r=R}\times
F_{B, m'}^{(2)}(k,r)\;;           &r < R\;,\\
M_{B,m'}^2 \big[F^{(2)}_{B, m'}(k,r)\big]'_{r=R}\times
G_{B, m'}^{(2)}(k,r)\;;       &r > R\;,
\end{array}\right.
\end{equation}

\begin{eqnarray}\label{d18}
\begin{array}{ll}
F_{B, m}^{(1)}(k,r) = {r}^{ 1/2 + |m|}e^{-\frac{1}{4}Br^2} {}_1F_1 \left({1\over 2} (|m| + m +1 -\frac{k-B}{B}), |m| +1; \frac{B}{2}r^2 \right),\\\\
G_{B, m}^{(1)}(k,r) ={r}^{ 1/2 +|m|} e^{-\frac{1}{4}Br^2} U \left({1\over 2} (|m| + m +1 -\frac{k-B}{B}), |m| +1; \frac{B}{2}r^2 \right)\;\;,
\end{array}
\end{eqnarray}

\begin{eqnarray}\label{d19}
\begin{array}{ll}
F_{B, m'}^{(2)}(k,r) = {r}^{ 1/2 + |m'|}e^{-\frac{1}{4}Br^2} {}_1F_1 \left({1\over 2} (|m'| + m' +1 -\frac{k+B}{B}), |m'| +1; \frac{B}{2}r^2 \right),\\\\
G_{B, m'}^{(2)}(k,r) ={r}^{ 1/2 +|m'|} e^{-\frac{1}{4}Br^2} U \left({1\over 2} (|m'| + m' +1 -\frac{k+B}{B}), |m'| +1; \frac{B}{2}r^2 \right)\;\;,
\end{array}
\end{eqnarray}

\begin{eqnarray}\label{d20}
M_{B,m}^1 = \bigg ( ||P_{B, m}^1 (k)||_{L^2 (]0, \infty))} \bigg )^{-1},
\end{eqnarray}

\begin{eqnarray}\label{d20}
M_{B,m'}^2 = \bigg ( ||P_{B, m'}^2 (k)||_{L^2 (]0, \infty))} \bigg )^{-1},
\end{eqnarray}

with

\begin{equation}\label{d21}
P_{B, m}^1(k,r)  = \left\{\begin{array}{ll}
 \big[G^{(1)}_{B,m}(k,r)\big]'_{r=R}\times
F_{B, m}^{(1)}(k,r)\;;           &r < R\;,\\
 \big[F^{(1)}_{B, m}(k, r)\big]'_{r=R}\times
G_{B, m}^{(1)}(k,r)\;;       &r > R\;,
\end{array}\right.
\end{equation}

and

\begin{equation}\label{15}
P_{B, m'}^2(k,r)  = \left\{\begin{array}{ll}
 \big[G^{(2)}_{B,m'}(k,r)\big]'_{r=R}\times
F_{B, m'}^{(2)}(k,r)\;;           &r < R\;,\\
\big[F^{(2)}_{B, m'}(k,r)\big]'_{r=R}\times
G_{B, m'}^{(2)}(k,r)\;;       &r > R\;.
\end{array}\right.
\end{equation}

The indicial equation also admits two solutions and $\dot H_{B , m,m'}$ has deficiency
indices $(2,2).$  Consequently, all self-adjoint (s.a)
extensions of $\dot H_{B , m,m'}$ are also given by a $4$-parameter
family of (s.a.) operators\cite{AG}. In particular, we define here

\begin{eqnarray}\label{122} H^{G_{m,m'}}_{B,m,m'}=
\left( \begin{array}{cl}
h_{B,m}^1 & \quad 0\\\\
0 &h_{B,m'}^2\end{array}
\right)
\end{eqnarray}

with the domain
\begin{eqnarray}\label{d24}
{\cal D}(H^{G_{m,m'}}_{B,m,m'}) & = & \left\{ f\in \bigg (L^2(]0, \infty [, dr)\cap H^{2,2}_{loc}(]0, \infty [-\{R\})\bigg) \bigotimes \C^2\, ;\right.
 f'(R_+)=f'(R_-)\equiv f'(R);\nonumber \\
& &f(R_+)-f(R_-) = G_{m,m'} f'(R); \left. \dot H_{B,m,m'}f \in L^2((0,\infty ))\bigotimes \C^2\, \,   \right\},\nonumber \\
& & m,\:m'\in {\Z}, \: G_{m,m'} \: \mbox{defined as previously}.
\end{eqnarray}

As expected, the continuity conditions are required here for the derivative function  $f'$ at the point $r=R$. This result respects the situation encountered in the case of the nonrelativistic $\delta '$ cylinder interaction.

The case $G_{m,m'} = O$ also coincides with the free kinetic energy Hamiltonian $H^{O}_{B,m,m'}$  for fixed quantum numbers $m,\:m'.$

Let $G=\{ G_{m,m'} \}_{m,\:m'\in \Z}$ and introduce in $L^2(\R^2) \bigotimes \C^2$ the operator

\begin{equation}\label{d25}
 H^{G}_{B}=\bigoplus_{m={-\infty}}^{m= +\infty} \bigoplus_{m'={-\infty}}^{m'= +\infty}U^{-1} H^{G_{m,m'}}_{B,m,m'}U\bigotimes 1\!\!1.
\end{equation}
By definition, $ H^{G}_{B}$ is the rigorous mathematical formulation of the formal expression $(\ref{d1}).$\\
Actually, it provides a slight generalization of $(\ref{d1})$, since $G$
may depend on $m, m'\in \Z.$\\

\smallskip\begin{thm}\label{2.1}
(i) The resolvent of $ H^{G_{m,m'}}_{B,m,m'}$ is given by

\begin{eqnarray}\label{d26}
{ (  H^{G_{m,m'}}_{B,m,m'} -k)}^{-1} = { ( H^{O}_{B,m,m'} -k)}^{-1} + \sum_{i,j =1}^{2}\mu_{i,j}(k)\left(\tilde \Psi_{B, m,m'}^i(\overline{k}), .\right)\tilde \Psi_{B, m,m'}^j(k),\nonumber \nonumber
\end{eqnarray}
\begin{equation}\label{20}
k\in \rho ( H^{G_{m,m'}}_{B,m,m'})\cap\rho ( H^{O}_{B,m,m'}) ,\:\:m,\:m'\in\Z,
\end{equation}
where
\begin{equation}\label{d27}
\mu_{1,1}(k) = \frac{\alpha_m}{M^{1}_{B,m}\bigg(G^{(1)}_{B, m}(k,R)F'^{(1)}_{B, m}(k,R) -G'^{(1)}_{B, m}(k,R)F^{(1)}_{B, m}(k,R) - \alpha_m G'^{(1)}_{B, m}(k,R)F'^{(1)}_{B, m}(k,R) \bigg)},
\end{equation}

\begin{equation}\label{d28}
\mu_{1,2}(k) = 0,
\end{equation}

\begin{equation}\label{d29}
\mu_{2,1}(k) = 0
\end{equation}

and
\begin{equation}\label{d30}
\mu_{2,2}(k) = \frac{\beta_{m'}}{M^{2}_{B,m'}\bigg(G^{(2)}_{B, m'}(k,R)F'^{(2)}_{B, m'}(k,R) -G'^{(2)}_{B, m'}(k,R)F^{(2)}_{B, m'}(k,R) - \beta_{m'} G'^{(2)}_{B, m'}(k,R)F'^{(2)}_{B, m'}(k,R) \bigg)}.
\end{equation}

The Green function $G_{B,m,m'}^{O}(k,r,r')$ of $H^{O}_{B,m,m' } $ has the form

\begin{equation}\label{d31} G_{B,m,m'}^{O}
 (k,r,r')=
\left( \begin{array}{ll}
g^{1,1}_{m, m'}(k,r,r')\qquad \qquad  \qquad& g^{1,2}_{m, m'}(k,r,r')\\\\   g^{2,1}_{m, m'}(k,r,r') \qquad \qquad  \qquad&  g^{2,2}_{m, m'}(k,r,r')\end{array}
\right).
\end{equation}

Since $G_{B,m,m'}^{O}(k,r,r')$ must satisfy the following equation:

\begin{equation}\label{d32} (H^{O}_{B,m,m' } -k)G_{B,m,m'}^{O}(k,r,r') =
\left( \begin{array}{ll}
\delta(r-r')\qquad \qquad  \qquad& 0\\\\  \qquad 0 &  \delta(r-r')\end{array}
\right),
\end{equation}

one has

\begin{eqnarray}\label{d33}
{(h^1_{B,m} -z)}g^{1,1}_{m, m'}(k,r,r') =\delta(r-r'),
\end{eqnarray}

\begin{eqnarray}\label{d34}
g^{1,2}_{m, m'}(k,r,r') = g^{2,1}_{m, m'}(k,r,r')=0 ,
\end{eqnarray}

\begin{eqnarray}\label{d35}
{(h_{B, m'}^2 -z)}g^{2,2}_{m, m'}(k,r,r')=\delta(r-r')
\end{eqnarray}

and we have

\begin{equation}\label{d36}
g^{1,1}_{m, m'}(k,r,r')= \left\{\begin{array}{ll}
M_{B,m}^1G'^{(1)}_{B, m}(k,r)\times
F_{B,m}^{(1)}(k,r')\;;           &r' < r\;,\\
M_{B,m}^1 F'^{(1)}_{B,m}(k,r)\times
G_{B,m}^{(1)}(k,r')\;;       &r' > r\;
\end{array}\right.
\end{equation}

and

\begin{equation}\label{d37}
g^{2,2}_{m, m'}(k,r,r')= \left\{\begin{array}{ll}
M_{B,m'}^2G'^{(2)}_{B, m'}(k,r)\times
F_{B,m'}^{(2)}(k,r')\;;           &r' < r\;,\\
 M_{B,m'}^2F'^{(2)}_{B,m'}(k,r)\times
G_{B,m'}^{(2)}(k,r')\;;       &r' > r\;.
\end{array}\right.
\end{equation}

We note that $g^{1,1}_{m, m'}(k,R,r)= \phi_{B, m,m'}^1(k,r)$ and $g^{2,2}_{m, m'}(k,R,r)= \phi_{B, m,m'}^2(k,r).$ \\
(ii) The resolvent of $H^{G}_{B}$ is given by

\begin{eqnarray}\label{d38}
{\left( H_{B}^G  - k^2\right)}^{-1}= {\left( H_{B}^O  - k^2\right)}^{-1} + \bigoplus_{m=-\infty}^{m=+\infty}  \bigoplus_{m'=-\infty}^{m'=+\infty} \sum_{i,j =1}^{2}\mu_{i,j}(k)\left(|.|^{-1}\tilde \Psi_{B, m,m'}^i(\overline{k}), .\right)|.|^{-1}\tilde \Psi_{B, m,m'}^j(k),\nonumber \nonumber
\end{eqnarray}
\begin{equation}\label{22}
k\in \rho ( H^{G}_{B})\cap\rho ( H^{O}_{B}) , \:\: m,\:m'\in\Z.
\end{equation}
\end{thm}\smallskip

{\it{Proof:}} Since $\dot H_{B,m,m'}$ has deficiency indices (2.2), it follows from Krein's formula \cite{AGHS} that the resolvent of $ H^{G_{m,m'}}_{B,m,m'}$ is given by

\begin{eqnarray}\label{d39}
{ ( H^{G_{m,m'}}_{B,m,m'} -k)}^{-1} = { ( H^{O}_{B,m,m'} -k)}^{-1} + \sum_{i,j =1}^{2}\mu_{i,j}(k)\left(\tilde \Psi_{B, m,m'}^i(\overline{k}), .\right)\tilde \Psi_{B, m,m'}^j(k),\nonumber \nonumber
\end{eqnarray}
\begin{equation}\label{23}
k\in \rho ( H^{G_{m,m'}}_{B,m,m'})\cap\rho ( H^{O}_{B,m,m'}),  \:\:m,\:m'\in\Z.
\end{equation}

For the determination of $\mu_{i,j}(k)$, we proceed as follows. Let $g\in L^2(]0,\infty])$ and define the function
 $$\chi_{m,m'}(k,r) = \left( {(  H^{G_{m,m'}}_{B,m,m'} -k)}^{-1}g\right) (r).$$
Since $\chi_{m,m'} \in D( H^{G_{m,m'}}_{B,m,m'})$, it follows that $\chi_{m,m'}$ should satisfy the boundary conditions of $D(H^{G_{m,m'}}_{B,m,m'})$. The implementation of these boundary conditions gives

\begin{equation}\label{d40}
\mu_{1,1}(k) = \frac{\alpha_m}{M^{1}_{B,m}\bigg(G^{(1)}_{B, m}(k,R)F'^{(1)}_{B, m}(k,R) -G'^{(1)}_{B, m}(k,R)F^{(1)}_{B, m}(k,R) - \alpha_m G'^{(1)}_{B, m}(k,R)F'^{(1)}_{B, m}(k,R) \bigg)},
\end{equation}

\begin{equation}\label{d41}
\mu_{1,2}(k) = 0,
\end{equation}

\begin{equation}\label{d42}
\mu_{2,1}(k) = 0
\end{equation}

and
\begin{equation}\label{d43}
\mu_{2,2}(k) = \frac{\beta_{m'}}{M^{2}_{B,m'}\bigg(G^{(2)}_{B, m'}(k,R)F'^{(2)}_{B, m'}(k,R) -G'^{(2)}_{B, m'}(k,R)F^{(2)}_{B, m'}(k,R) - \beta_{m'} G'^{(2)}_{B, m'}(k,R)F'^{(2)}_{B, m'}(k,R) \bigg)}.
\end{equation}

Inserting (\ref{d40}), (\ref{d41}),  (\ref{d42}) and (\ref{d43})                     into (\ref{d39}), we deduce the expression  (\ref{d26}).
Eq. (\ref{d38}) follows from (\ref{d25}) and (\ref{d39}) $.$\\

Spectral properties of $ H^{G_{m,m'}}_{B,m,m'}$ are provided by the following theorem.

\smallskip\begin{thm}\label{2.3}:
For all $\alpha_m\in (-\infty ,\infty )$ and $\beta_{m'}\in (-\infty ,\infty ),$ we have the following
results
\begin{eqnarray}\label{sppp}
\sigma_{ess}( H^{G_{m,m'}}_{B,m,m'}) = \emptyset,
\end{eqnarray}

\begin{eqnarray}\label{s}
\sigma_{sc}( H^{G_{m,m'}}_{B,m,m'})= \emptyset
\end{eqnarray}
\begin{eqnarray}\label{sp30}
\sigma_p(H^{G_{m,m'}}_{B,m,m'}) & =&\bigg \{E\in \R/ G'^{(1)}_{B, m}(E,R)F^{(1)}_{B, m}(E,R) -G^{(1)}_{B, m}(E,R)F'^{(1)}_{B, m}(E,R)\nonumber\\
& & - \alpha_m G'^{(1)}_{B, m}(E,R)F'^{(1)}_{B, m}(E,R) =0 \nonumber\\
& & \mbox{or}\:\:  G'^{(2)}_{B, m'}(E,R)F^{(2)}_{B, m'}(E,R) -G^{(2)}_{B, m'}(E,R)F'^{(2)}_{B, m'}(E,R)\nonumber\\
& & - \beta_{m'} G'^{(2)}_{B, m'}(E,R)F'^{(2)}_{B, m'}(E,R) =0    \bigg\}.\nonumber \\
\end{eqnarray}
The negative eigenvalues of $ H^{G_{m,m'}}_{B,m,m'}$ are obtained from the equation
\begin{equation}\label{a40}
{ det(\mu_{i,j}(E))}^{-1} =   0;\:\:E<0,
\end{equation}
which has at most two solutions $E_0 <0.$
\end{thm}\smallskip

{\it{Proof: }}

Similar to the proof of Theorem 2.2$.$

 \section{Remarks}
Let us point out that the nonrelativistic results can be trivially
 deduced from the relativistic one setting the scalar parameter $\beta =0.$

 The results for the point interaction at the origin appear as a particular case of the cylinder interaction investigated here. Indeed, when $R \rightarrow 0$ in (\ref{c1}) and (\ref{d1}), we recover the boundary conditions  corresponding to the nonrelativistic point interaction investigated in \cite{GHS}.  Finally, let us recall that the properties for point interaction placed at any point $x$ could be found using the transformation relation $t_x H_\alpha t_{-x}$ given in \cite{AGHS}, where $t_x$ is a translation application of vector $x,$ $H_\alpha$ being the Hamiltonian perturbed by point interaction at the origin $r=0$.
\\
\\

{\bf{Acknowledgments:}} The authors thank the Belgian cooperation CUD - CIUF/UAC - IMSP,
 the Abdus Salam International Center for Theoretical Physics and the Conseil Regional Provence - Alpes - Cote d'Azur (France) for their financial support.

\end{document}